\documentclass[a4paper,journal,onecolumn,draftclsnofoot,11pt]{IEEEtran}
\usepackage{cite}

\usepackage{amsmath,amssymb,subfigure,amsthm,mathtools,enumerate,setspace}
\usepackage{graphicx}
\usepackage{color}

\newcommand{\A}{\mathcal{A}}

\newcommand{\R}{\mathbb{R}}

\newcommand{\Z}{\mathbb{Z}}
\newcommand{\N}{\mathbb{N}}

\newcommand{\qnl}{ {q^{(\lambda)}_n}}
\newcommand{\ql}{ {q^{(\lambda)}}}
\newcommand{\Ql}{\tilde{q}^{(\lambda')}}
\newcommand{\Qnl}{\tilde{q}^{(\lambda')}_n}

\newcommand{\fl}{f^{(\lambda)}}
\newcommand{\fnl}{\big(\fl\big)_n}
\newcommand{\tgl}{\tilde{g}_{t}^{[\lambda']}}
\newcommand{\tgln}{\big(\tilde{g}_t^{[\lambda']}\big)_n}
\newcommand{\sd}{\Sigma\Delta}
\newcommand{\bnd}{\lambda'\pi}

\newcommand{\be}{\begin{enumerate}}
\newcommand{\bi}{\begin{itemize}}
\newcommand{\ee}{\end{enumerate}}
\newcommand{\ei}{\end{itemize}}

\newtheorem{theorem}{Theorem}[]
\newtheorem{lemma}[theorem]{Lemma}

\theoremstyle{remark}
\newtheorem{remark}{Remark}[theorem]
\theoremstyle{definition}

\begin{document}
%
\onehalfspace
\title{A Deterministic Analysis of Decimation for Sigma-Delta Quantization of Bandlimited
 Functions}




\author{ Ingrid Daubechies
and  Rayan Saab
\thanks{I. Daubechies is with the Department of Mathematics, Duke University.} 
\thanks{R. Saab is with the Department of Mathematics, University of California San Diego}}

\maketitle

\begin{abstract}
We study Sigma-Delta ($\sd$) quantization of oversampled bandlimited functions. We prove that digitally integrating blocks of bits and then down-sampling, a process known as decimation, can efficiently encode the associated $\sd$ bit-stream. It allows a large reduction in the bit-rate while  still permitting good approximation of the underlying bandlimited function via an appropriate reconstruction kernel.  Specifically, in the case of stable $r$th order $\sd$ schemes we show that the reconstruction error decays exponentially in the bit-rate. For example, this result applies to the 1-bit, greedy, first-order $\sd$ scheme.
\end{abstract}

\section{Introduction}

Analog-to-digital (A/D) conversion is the process by which signals (viewed as vectors) are represented by bit streams to allow for digital storage, transmission, and processing using modern computers. Typically, A/D conversion is thought of as being composed of sampling and quantization.   Sampling consists of collecting inner products of the signal with appropriate vectors. Quantization consists of replacing these inner products with elements from a finite set, known as the quantization alphabet. Often, quantization is followed by  some form of encoding or compression, in order to reduce the size or bit-rate of the digital data. A good A/D scheme allows for accurate reconstruction of the original object from its quantized (and compressed) samples. 
Sigma-Delta ($\sd$) quantization was proposed in the 1960's \cite{inose1963unity} as a method for digitizing bandlimited functions. In fact, $\sd$ quantization  schemes remain in use today, in large part due to their robustness to errors caused by circuit imperfections, but also due to their ability to trade-off quantizer bit-depth and oversampling (cf. \cite{daub-dev}). 

In the context of bandlimited functions, oversampling ---coupled with an appropriate $\sd$ quantization scheme--- enables one to use coarse (even binary) quantization alphabets, such as $\mathcal{A}:=\{\pm 1\}$, and then to reconstruct the function accurately from the resultant bit-stream. In particular, $\sd$ schemes have been devised \cite{G-exp, DGK10} whereby the reconstruction error, measured in the $L^\infty$ norm, decays exponentially fast in the oversampling rate. Specifically,  \cite{G-exp} and \cite{DGK10} each devise a family of sophisticated $\sd$ schemes parametrized by an order $r$, and choose an appropriate scheme (from this family) by optimizing $r$ as a function of the oversampling rate. Working with the alphabet $\A=\{\pm 1\} $, and denoting the oversampling rate by $\lambda$, the best known reconstruction error guarantees (see \cite{DGK10}) behave like $2^{-c\lambda}$, with $c\approx 0.1$. In this context, since the size of the alphabet is fixed, the bit-rate resulting from $\sd$ quantization is proportional to the oversampling rate. 
Consequently, the reconstruction error of \cite{G-exp} and \cite{DGK10}  decays exponentially fast with the bit-rate, albeit with a sub-optimal coefficient in the exponent.\footnote{For example, given a bit-rate of $\lambda$ bits per Nyquist interval, one can obtain exponential decay in $\lambda$ (with a much better, essentially optimal (see, e.g., \cite{G-exp}), coefficient in the exponent) by sampling at slightly higher than the Nyquist rate and replacing the samples by their binary approximations. In particular, the $L^\infty$ error is $O(2^{-\lambda})$. On the other hand, this method is not robust to errors in assigning the bits (cf. \cite{daub-dev}).} 

In this note, we prove that using any stable $r$th order $\sd$ schemes, with an arbitrary integer $r>0$ (including the $1st$ order, greedy, $\sd$ scheme defined below) followed by a simple encoding step, we can always reconstruct a bandlimited function from its encoded bit-stream with a reconstruction error that decays exponentially fast in the bit-rate. Moreover, we obtain a near-optimal coefficient in the exponent.

\subsection{Preliminaries}
We define the Fourier transform, $\hat{f}$, of $f\in L^2(\R)$ via
$$\hat{f}(\omega) ~=~ \frac{1}{\sqrt{2\pi}} \int\limits_{-\infty}^{\infty} f(t)e^{-i\omega t}dt.$$
The inverse Fourier transform is then given by
$${f}(t) ~=~ \frac{1}{\sqrt{2\pi}} \int\limits_{-\infty}^{\infty} \hat{f}(\omega)e^{i\omega t}d\omega.$$
In this note we are interested in bandlimited functions $f\in L^2(\R)$ with $|f(t)|<1$ and with Fourier transform vanishing outside the interval $[-\pi , \pi].$ We denote the set of such functions by $B_\pi$.  
The classical sampling theorem yields a method of reconstructing an arbitrary $f\in B_\pi$ perfectly from its so-called Nyquist rate samples $f(n), n\in\Z$. In particular,  \begin{equation}f(t)=\sum\limits_{n\in\Z}f(n)\frac{\sin(\pi(t-n))}{\pi(t-n)}.\label{eq:Shannon}\end{equation}
Nevertheless, sampling at this Nyquist rate is rarely done in practice because the reconstruction kernel $\frac{\sin(\pi(t-n))}{\pi(t-n)}$ decays too slowly. This implies that if one were to reconstruct with ``noisy" samples $f(n)+\varepsilon_n$ (instead of with $f(n)$ in \eqref{eq:Shannon})  large, possibly unbounded, reconstruction errors could result, even if $\epsilon_n$ were bounded. This makes \eqref{eq:Shannon} unsuitable for reconstruction from quantized samples. Instead, one may revert to {\emph{oversampling}}, i.e.,  collect the samples $f(n/\lambda)$ for some $\lambda>1$ and then reconstruct via the formula 
\begin{equation} f(t)=\frac{1}{\lambda}\sum\limits_{n\in\Z}f(n/\lambda)g(t-n/\lambda), \label{eq:rec_formula}\end{equation}
where $g$ is a function with $\hat{g} \in C^{\infty}$, $\hat{g}(\omega)=\frac{1}{\sqrt{2\pi}}$ for $|\omega| \leq \pi$ and $\hat{g}(\omega)=0$ for $|\omega|\geq \lambda\pi.$
With these sampling and reconstruction schemes, it can be seen (cf. \cite{daub-dev}) that the reconstruction error induced by small errors in the sample values is small. In the worst case, it is proportional to the error in the samples. On the other hand, in the quantization setting one has control over how the samples $f(n/\lambda)$ are replaced by elements from $\mathcal{A}$, so one can do significantly better. 

\subsection{$\sd$ quantization and prior work}
One-bit, first order, greedy $\sd$ quantization produces bits ${ \qnl} \in \{ -1, 1 \}$ via\footnote{Here and throughout, we use the superscript $\lambda$ to indicate the oversampling rate at which a discrete sequence is obtained.} the following recursion, with initial condition $u_0, |u_0| < 1$:  
\begin{equation}
\qnl = \textrm{sign} \left( u_{n-1}+ f(n/\lambda)  \right), 
\label{equ:FirstOrdq}
\end{equation}
\begin{equation}
u_n =   u_{n-1} + f(n/\lambda)  - \qnl.
\label{equ:FirstOrdu}
\end{equation}
One can see, by induction, that $|u_n| < 1$ for all $n$.
Moreover, using this scheme for quantization and the function $g$ in \eqref{eq:rec_formula} for reconstruction, we have (see \cite{daub-dev})  $$|f(t) - \sum_{n\in{\Z}}\qnl g(t-n/\lambda) | \leq C_g/\lambda.$$
To generalize the above $\sd$ scheme, let $r$  be a positive integer and denote by $\mu: \R^{r+1} \to \R$ the  ``quantization rule". One can then define an $r$th order $\sd$ scheme via the recursion:
\begin{equation}
\qnl = \textrm{sign}\left(\mu(f(n/\lambda),u_{n-1},u_{n-2},\dots,u_{n-r})\right),
\label{equ:rthOrdq}
\end{equation}

\begin{equation}
(\Delta^r u)_n =  f(n/\lambda) - \qnl,
\label{equ:rthOrdu}
\end{equation}
where the operation of the difference operator $\Delta$ on a sequence $h$ is 
defined by $(\Delta h)_n:= h_n-h_{n-1}$; (\ref{equ:rthOrdu}) is
equivalent to

\begin{equation}
u_n =  f(n/\lambda) - \qnl - \sum^{r}_{j=1} {r \choose j} (-1)^j u_{n-j}.
\label{equ:rthOrdu_v2}
\end{equation}
An important issue in the design and analysis of higher order schemes is ensuring that the sequence $u_n$ is uniformly bounded via a proper choice of $\mu$. 
Thus, we say that an $r$th order $\sd$ scheme is stable if $\|u\|_\infty \leq  C_{\sd}$ whenever $|f(n/\lambda)|<1$   for some constant $C_{\sd}$ that may depend on $r$.  Daubechies and DeVore \cite{daub-dev} proposed the first family of stable $\sd$ quantization algorithms and used them to obtain error bounds of the form 
$$|f(t) - \sum_{n\in{\Z}}\qnl g(t-n/\lambda) | \leq C(r)/\lambda^r.$$
By choosing the optimal $r(\lambda)$, they also derived the improved estimate 
$$|f(t) - \sum_{n\in{\Z}}\qnl g(t-n/\lambda) | \leq \tilde{C}\lambda^{-c \log{\lambda}}.$$
G{\"u}nt{\"u}rk \cite{G-exp} proposed a different family of $\sd$ schemes and used them to obtain the bound 
$$|f(t) - \sum_{n\in{\Z}}\qnl g(t-n/\lambda) | \lesssim 2^{-c\lambda},$$
with $c \approx 0.07$, again by choosing the order $r$ as a function of $\lambda$. Deift et al. \cite{DGK10} improved this result by obtaining the coefficient $c\approx 0.102$ in the exponent. 

For the case of constant input to the $\sd$ quantization, there has been some work (cf. \cite{hein1992new, gunturk2001robustness, ayaz2009sigma}) seeking upper bounds on the number of possible $\sd$ bit-sequences of length $N$. For example \cite{hein1992new} showed that asymptotically, for first-order $\sd$ schemes, the number of such sequences is $O(N^2)$. These sequences can be represented by binary labels of length $O(\log(N))$ while still enabling a reconstruction error of $1/N$. However, no analogous bound is known for bandlimited functions. 

In practice, when working with oversampled A/D conversion of bandlimited functions, it is common to incorporate a so-called decimation step (see, e.g., \cite{Candy86}). This process reduces the bit-rate by mapping blocks of quantized samples (obtained at a high oversampling rate) to elements from a codebook (another finite set). 
An analysis of such techniques was given by Candy \cite{Candy86}, under the simplifying (albeit generally false) assumption that $\sd$ quantization introduces random ``noise" that is uncorrolated with the input. The conclusion of \cite{Candy86}, based on the randomness assumption and numerical experiments, is that decimation can produce dramatic decreases in the bit-rate without compromising the quality of approximation. In this note, we provide a rigorous mathematical analysis of decimation, with the same conclusion.  

\section{Main result}
We prove that by digitally integrating blocks of bits produced by one bit, $r$th order, stable $\sd$ schemes ---a process known in the engineering community as decimation \cite{Candy86}--- we can reduce the number of bits per Nyquist interval from $\lambda$ to approximately $ r\log{\lambda}$. We prove that this still allows for an approximation error that decays like $1/\lambda^r$, albeit via a different reconstruction kernel $\tilde{g}$ than that of \eqref{eq:rec_formula}. In other words, we show exponential decay of the approximation error as a function of the bit-rate, with a near-optimal exponent.

To make the discussion more concrete, let us start with some definitions. For a sequence $h$, and positive integers $r,\,\rho \geq 1$, define the $r$th order partial sums
\begin{align*}
 (S^{r}_\rho h)_n&:=\frac{1}{2\rho+1}\sum\limits_{m=-\rho}^{\rho}(S^{r-1}_\rho h)_{n-m}
\\ &=\frac{1}{2\rho+1}\sum\limits_{m=n-\rho}^{n+\rho}(S^{r-1}_\rho h)_{m},
 \end{align*}
where  $(S^{0}_\rho h)_n:=h_n.$ For a bit-sequence $\ql$ generated from an $r$th order $\sd$ quantization of a bandlimited function, and for an integer $\rho < \frac{\lambda-1}{2}$, we are interested in the  integrated bit sequence $(S^{r}_\rho \ql)_n$, as well as its decimated (subsampled) version
\[\Qnl:=\Big(S^{r}_\rho \ql\Big)_{(2\rho+1)n}.\]
We prove the following theorem.
\begin{theorem}\label{thm:I_intro}
Suppose that $f$ is in $B_\pi$ , $\rho\in \N\cap (1,\frac{\lambda-1}{2})$,  and define $\lambda':= \frac{\lambda}{2\rho+1}$. Then the following are true of 1-bit stable $r$th order $\sd$ quantization. 
 \be \item[(i)] There exists a function $\tilde{g}$ such that 
$$ \left| \frac{1}{\lambda'} \sum\limits_{n\in\Z} \Qnl \tilde{g}(t-n/\lambda') - f(t) \right| \leq  C_{\sd}C^r\Big(\frac{\lambda'}{\lambda}\Big)^r=:\mathcal{D}.$$
\item[(ii)] 
To encode $\Qnl$, one needs at most $\lambda' \log_2\Big((2\rho+1)^r+1\Big)$ bits per Nyquist interval where \begin{align*}\lambda' \log_2\big((2\rho+1)^r+1\big) 
 &\leq  \lambda' \log_2{(2\Big(\frac{\lambda}{\lambda'}\Big)^r)}=:\mathcal{R}.\end{align*}
\ee
Consequently 
\begin{equation} \label{eq:RD}
 \mathcal{D}(\mathcal{R})=2C_{\sd}C^r   
 2^{-{\mathcal{R}}/{\lambda'}}.
 \end{equation}
 Here $C>1$ is a constant independent of $\lambda$, $\lambda'$ and $r$. $C_{\sd}$ is a constant that depends on the scheme (i.e., possibly on $r$).
\end{theorem}

\begin{remark}
As $\lambda$ grows, we may select a progressively larger $\rho$, so that in the limit $\lambda'$ approaches 1. Hence the claim about near-optimality. 
\end{remark}

\begin{remark}
Examining the proof of the theorem (below), one should be able to extend the proof without too much difficulty to the case of multi-bit quantization. For ease of exposition, we refrain from doing this in this note.  
\end{remark}

\section{Proof of Theorem \ref{thm:I_intro}}
\proof We will begin by proving (i). Our goal is to bound the error \begin{equation} e :=  \Big| \frac{1}{\lambda'} \sum\limits_{n\in\Z} \Qnl \tilde{g}(t-n/\lambda') - f(t)\Big|. \end{equation} 
To that end, let us first define the sequence $\fnl:=f(n/\lambda)$.  Using the triangle inequality, we have $e \leq e_1 + e_2$
where \begin{align*}e_1 &:= \Big | \frac{1}{\lambda'} \sum\limits_{n\in\Z} \Big(\Qnl -(S^{r}_\rho \fl)_{(2\rho+1)n}\Big)\tilde{g}(t-n/\lambda') \Big| \\ &=  \Big | \frac{1}{\lambda'} \sum\limits_{n\in\Z} \Big(S^{r}_\rho (\ql - \fl)\Big)_{(2\rho+1)n}\cdot\tilde{g}(t-n/\lambda') \Big|  \end{align*}

and \[e_2 := \Big|\frac{1}{\lambda'} \sum\limits_{n\in\Z} \big(S^r_\rho  \fl \big)_{(2\rho+1)n}\cdot \tilde{g}(t-n/\lambda') -f(t)\Big|.\]
The remainder of the proof will consist of bounding $e_1$ and showing that there exists a function $\tilde{g}$ for which $e_2=0$. Along the way we will specify $\tilde{g}$.

To bound $e_1$, we first define, for an integer $p$, the difference operators $\Delta_p$, and $\bar{\Delta}_p$ by their actions $\Big(\Delta_p x\Big)_n: =  x_{n+p}-x_{n-p-1}$  and  $\Big(\bar\Delta_p x\Big)_n: =  x_{n}-x_{n+2p+1}$, respectively. 
One easily checks that $S^{1}_\rho \Delta = \frac{1}{2 \rho + 1} \Delta_\rho$
and similarly $S^{r}_\rho \Delta^r = \frac{1}{(2 \rho + 1)^r} \Delta^r_\rho$,
where $\Delta^r_\rho := (\Delta_\rho)^r $.
For convenience, we introduce the notation $\tgln := \tilde{g}(t-n/\lambda')$ and observe that $\tgln = (\tilde{g}_t^{[\lambda]})_{(2\rho+1)n}$. Using the $\sd$ state equations \eqref{equ:rthOrdq}, \eqref{equ:rthOrdu}, and then reindexing we can write
\begin{align}
e_1 	&=\Big| \frac{1}{\lambda'}\left(\frac{1}{2\rho +1}\right)^r\sum\limits_{n\in\Z} (\Delta_\rho^r u)_{(2\rho+1)n}\tgln  \Big|\\
	&=\Big| \frac{1}{\lambda'}\left(\frac{\lambda'}{\lambda}\right)^r \sum\limits_{n\in \Z}u_{n(2\rho+1)+\rho}  (\bar\Delta^r_0 \tgl)_n   \Big|	\\
	&\leq \frac{1}{\lambda'}\left(\frac{\lambda'}{\lambda}\right)^r \cdot \Big(\frac{1}{\lambda'}\Big)^{r-1} \|\tilde{g}^{(r)}\|_{L_1}\|u\|_\infty 
	 = \frac{1}{\lambda^r}\|\tilde{g}^{(r)}\|_{L_1}\|u\|_\infty.\label{eq:RHS}
\end{align}
The last inequality is due to (the proof of) Proposition 3.1 in \cite{daub-dev};
the notation $\tilde{g}^{(r)}$ stands here for the $r$th derivative of
the function $\tilde{g}$. We shall now turn to controlling $e_2$, and return to the right hand side of \eqref{eq:RHS} shortly. 

To bound  $e_2$, let us first extend the use of our notation for partial sums so that for integers $r\geq 1$, $(S^r{f})(t):=\frac{1}{2\rho+1}\sum\limits_{m=-\rho}^{\rho}(S^{r-1}f)(t-m/\lambda)$ where $(S^0f)(t)=f(t)$. Thus, taking Fourier transforms 
\begin{equation}
\widehat{(S^r{f})}(\omega) = \Big(\frac{\sin(\frac{2\rho+1}{2\lambda}\omega)}{(2\rho+1)\sin(\frac{1}{2\lambda}\omega)}\Big)^r \hat{f}(\omega)
=\Big(\frac{\sin(\frac{1}{2\lambda'}\omega)}{\frac{\lambda'}{\lambda}\,\sin(\frac{1}{2\lambda}\omega)}\Big)^r \hat{f}(\omega).
\end{equation}
Let $\hat{h}(\omega)\in C^\infty$ satisfy

\begin{align} \hat{h}(\omega):= \left\{\begin{array}{ccc} 
1, & ~ \omega =0\\
\frac{\lambda\sin(\frac{1}{2\lambda}\omega)}{\lambda' \sin(\frac{1}{2\lambda'}\omega)}, & ~|\omega|\leq \pi \\  0, & |\omega|\geq \lambda'\pi.\end{array}\right. \end{align} 
Since $\hat{f}$ is compactly supported, using Fourier series we have
\begin{equation}
\widehat{(S^r{f})}(\omega) = \sum_{n\in\Z} c_n e^{-i\omega n/\lambda'}\cdot \hat{g}(\omega).
\end{equation}
where $\hat{g} \in C^{\infty}$, $\hat{g}(\omega)=\frac{1}{\sqrt{2\pi}}$ for $|\omega| \leq \pi$, $\hat{g}(\omega)=0$ for $|\omega|\geq \lambda'\pi.$ Here,
\[c_n = \frac{1}{\sqrt{2\pi} \lambda'}\int\limits_{-{\pi}{\lambda'}}^{\pi\lambda'}\widehat{(S^r{f})}(\omega)e^{i\omega n/ \lambda'}d\omega = \frac{1}{\lambda'}{(S^r{f})}(n/\lambda').\]
Thus, we deduce that
\begin{align}
f(t)	&= \frac{1}{\sqrt{2\pi}} \int\limits_{-\pi \lambda}^{\pi \lambda} \hat{f}(\omega)e^{i\omega t}d\omega 
	= \frac{1}{\sqrt{2\pi}} \int\limits_{-\pi \lambda}^{\pi \lambda} \widehat{(S^r{f})}(\omega) \hat{h}(\omega)^r e^{i\omega t}d\omega \\
	&= \frac{1}{\sqrt{2\pi}\lambda'} \int\limits_{-\pi \lambda}^{\pi \lambda} \sum_{n\in\Z} {(S^rf)}(n/\lambda') e^{-i\omega n/\lambda'}\cdot \hat{h}(\omega)^r\hat{g}(\omega) e^{i\omega t}d\omega\\
	&= \frac{1}{ \lambda'}  \sum_{n\in\Z} {(S^r{f})}(n/\lambda') \frac{1}{\sqrt{2\pi}} \cdot \int\limits_{-\pi \lambda}^{\pi \lambda}  \hat{h}(\omega)^r\hat{g}(\omega)e^{i\omega (t-n/\lambda')}d\omega.
	\end{align}
Let $h_r(t)$ be the inverse Fourier transform of $\hat{h}^r(\omega)$ and denote by $\tilde{g}(t):= (g * h_r)(t)$ the convolution of $g$ and $h_r$. We now have that $f(t)= \frac{1}{\lambda'}  \sum_{n\in\Z} {(S^r{f})}(n/\lambda') \tilde{g}(t-n/\lambda'),$ i.e., that $e_2=0$. To conclude the proof of (i), we note that $\|\tilde{g}^{(r)}\|_{L_1} = \|g^{(r)}*h_r \|_{L_1} \leq \|g^{(r)}\|_{L_1} \| h_r \|_{L_1} \leq  C^r \lambda'^r$ where the last inequality is a direct consequence of Lemma \ref{lem:helper} below and the fact that $\|g^{(r)}\|_{L_1}$ can be treated as a constant. Noting that $\|u\|_\infty \leq C_{\sd}$ completes the proof.

To prove (ii), note that the sum of $2\rho+1$ elements each taking on values in $\{\pm 1\}$, is an odd integer in $[-(2\rho+1) , 2\rho+1 ]$. There are $2\rho+2$ such integers, so each element of the sequence $S^{1}\ql$ 
can be encoded using $\log_2(2\rho+2)$ bits. Similarly, the sum of $2\rho+1$ odd integers in $[-(2\rho+1) , 2\rho+1 ]$, is an odd integer in $[-(2\rho+1)^2 , (2\rho+1)^2]$. There are $(2\rho+1)^2+1$ such integers. Proceeding in this fashion, we see that each $\Qnl$ can be encoded using $\log_2\big((2\rho+1)^r+1\big)$ bits.  Moreover, note that due to decimation, for every $\lambda$ original $\sd$ bits of $\ql$ there are $\lambda'=\frac{\lambda}{2\rho+1}$ elements of $\Ql$. The rate-distortion relationship then follows by combining (i) and (ii). 
\qed
%
\appendix
\begin{lemma} \label{lem:helper}Let $\hat{\phi}_0(\omega)$ be in $C^\infty$ and bounded, with  $\hat{\phi}_0(\omega)=1$ when $|\omega|\leq 1$ and $\hat{\phi}_0(\omega)=0$ when $|\omega|\geq c$ for some fixed $c\in (1,\infty)$. Define $\hat{\phi}(\omega):=\hat{\phi}_0\big(\frac{\omega}{\pi}\big)$. Let $\lambda'>c$, let  $\hat{h}_0(\omega)=\frac{\lambda\sin(\frac{\omega}{2\lambda})}{\lambda' \sin(\frac{\omega}{2\lambda'})},$  and define
 \begin{align}\hat{h}(\omega):= \hat{h}_0(\omega) \hat{\phi}(\omega)=\hat{h}_0(\omega) \hat{\phi}_0(\omega/\pi).\label{eq:hat_h}\end{align}  Then $\|h\|_{L_1}=\int_{-\infty}^{\infty} |h(t)|dt \leq C\lambda'$ where $C$ depends on $\phi_0.$
 Consequently, for any $r\geq1$, denoting by $h_r(t)$ the inverse Fourier transform of $\hat{h}^r(\omega)$, we have $\|h_r\|_{L_1}\leq C^r \lambda'^r.$
\end{lemma}
\proof
Note that 
\begin{align}\int_{-\infty}^{\infty} |h(t)|dt &= \frac{1}{\sqrt{2\pi}}\int_{-\infty}^{\infty} \frac{1}{t^2+1} \Big|\int_{-\bnd}^{\bnd} \hat{h}(\omega)e^{i \omega t}d\omega \Big|dt \nonumber\\ & \quad+  \frac{1}{\sqrt{2\pi}}\int_{-\infty}^{\infty} \frac{1}{t^2+1} \Big|\int_{-\bnd}^{\bnd}t^2 \hat{h}(\omega)e^{i \omega t}d\omega \Big|dt.\label{eq:split}
\end{align}
We will proceed by bounding each of the summands on the right hand side separately.  The first term is controlled by 
\begin{align}
 \frac{1}{\sqrt{2\pi}} & \int_{-\infty}^{\infty} \frac{1}{t^2+1} \Big|\int_{-\bnd}^{\bnd} \hat{h}(\omega)e^{i \omega t}d\omega \Big|dt \nonumber\\ &\leq  \frac{1}{\sqrt{2\pi}}\int_{-\infty}^{\infty} \frac{1}{t^2+1} \lambda' \pi^2 C_{\phi_0}  dt\nonumber
\leq \frac{\pi^{5/2}}{\sqrt{2}}C_{\phi_0} \lambda',
\label{eq:term1}
\end{align}
where $C_{\phi_0}= \sup\limits_\omega|\hat\phi_0(\omega)|$ and the first inequality is due to the bound $|\hat{h}_0(\omega)|\leq \pi/2$ when $|\omega|\leq \lambda'\pi.$
To control the second term, we observe that 
\begin{align}
& \frac{1}{\sqrt{2\pi}}  \int_{-\infty}^{\infty} \frac{1}{t^2+1} \Big|\int_{-\bnd}^{\bnd}t^2 \hat{h}(\omega)e^{i \omega t}d\omega \Big|dt \nonumber \\
 &=  \frac{1}{\sqrt{2\pi}}\int_{-\infty}^{\infty} \frac{1}{t^2+1} \Big|\int_{-\bnd}^{\bnd} \hat{h}(\omega)(e^{i \omega t})''d\omega \Big|dt \nonumber \\
&\leq \frac{1}{\sqrt{2\pi}}\int_{-\infty}^{\infty} \frac{1}{t^2+1} \Big| \int_{-\bnd}^{\bnd} \hat{h}''(\omega)(e^{i \omega t})d\omega - \hat{h}'(\omega)e^{i\omega t}|_{-\bnd}^{\bnd}\Big|dt \nonumber\\
&\leq \frac{1}{\sqrt{2\pi}}\int_{-\infty}^{\infty} \frac{1}{t^2+1} \Big( \int_{-\bnd}^{\bnd} |\hat{h}''(\omega)|d\omega +\Big| \hat{h}'(\omega)e^{i\omega t}|_{-\bnd}^{\bnd}\Big|\Big) dt.
\end{align}

Above, the first inequality is due to integration by parts. In particular, 
\begin{align*} \int_{-\bnd}^{\bnd}  &\hat{h}(\omega)(e^{i \omega t})''d\omega 
=  \hat{h}(\omega)(e^{i \omega t})' |_{-\bnd}^{\bnd} - \int_{-\bnd}^{\bnd} \hat{h}'(\omega)(e^{i \omega t})'d\omega  \\
&= 0 - \hat{h}'(\omega)e^{i\omega t}|_{-\bnd}^{\bnd}+ \int_{-\bnd}^{\bnd} \hat{h}''(\omega)(e^{i \omega t})d\omega.
\end{align*}
Thus
\begin{align}
 \frac{1}{\sqrt{2\pi}}&\int_{-\infty}^{\infty} \frac{1}{t^2+1} \Big|\int_{-\bnd}^{\bnd}t^2 \hat{h}(\omega)e^{i \omega t}d\omega \Big|dt \\
&\leq \frac{1}{\sqrt{2\pi}}\int_{-\infty}^{\infty} \frac{1}{t^2+1} \cdot (2\bnd C_{\hat{h}''}+2C_{\hat{h}'})dt \nonumber\\
&\leq \sqrt{2\pi} (\bnd C_{\hat{h}''}+C_{\hat{h}'}),\label{eq:term2}
\end{align}
where the constants satisfy $C_{\hat{h}''}\geq |\hat{h}''(\omega)|$ for all $\omega \in [-\bnd , \bnd]$ and $C_{\hat{h}'} \geq |\hat{h}'(\omega)|$ for all $\omega \in [-\bnd , \bnd]$. 

To compute $C_{\hat{h}'}$, we observe that the function ${\hat{h}_0}'(\omega):= \frac{ \lambda'  \sin(\frac{\omega}{2\lambda'})\cos(\frac{\omega}{2\lambda}) -\lambda  \sin(\frac{\omega}{2\lambda})\cos(\frac{\omega}{2\lambda'}) }{2\lambda'^2 \sin^2(\frac{\omega}{2\lambda'})}$ achieves its maximum absolute magnitude  on $[-\lambda'\pi, \lambda'\pi]$ at $\pm \lambda'\pi$. Denoting this maximum by $C_1$, we have 
$C_1=\frac{\cos{\frac{\lambda'\pi}{2\lambda}}}{2\lambda'} \leq \frac{1}{2}$,
since $\lambda'>1$.
Similarly, one can verify that $\hat{h}_0''(\omega)$ achieves its maximum amplitude on $[-\lambda'\pi, \lambda'\pi]$ at $\pm \lambda'\pi$. A simple evaluation then reveals that the maximum, denoted by $C_2$, is $\frac{ \sin({\frac{\pi\lambda'}{2\lambda}}) (\lambda^2-\lambda'^2)}{4\lambda \lambda'^3}$. Using that $\sin(x)\leq x$,  we observe that $C_2 \leq \pi/8$.
Next, observe that 
$\hat{\phi}(\omega)=\hat{\phi}_0(\omega/\bnd)$ 
thus 
 \begin{align}|
 \hat{h}'(\omega)|&=|(\hat{h}_0(\omega)\hat{\phi}(\omega))'|\nonumber
 \\ &\leq |\hat{h}_0'(\omega)\hat\phi_0(\omega/\pi)|+|\hat{h}_0(\omega)\hat{\phi}_0'(\omega/\pi)/\pi|\nonumber
 \\ &\leq \frac{1}{2}C_{\phi_0}+\frac{\pi}{2}\cdot \frac{C_{\phi_0'}}{\pi},\nonumber
 \end{align}
where $C_{\phi_0'} = \sup\limits_\omega |\hat\phi'_0(\omega)| $. 
Similarly, 
 \begin{align}
 %
  |\hat{h}''(\omega)|&=|(\hat{h}_0(\omega)\hat{\phi}(\omega))''| \nonumber
 \\ &\leq |\hat{h}_0''(\omega)\hat{\phi}_0(\omega/\pi)|+|2\hat{h}'_0(\omega)\hat{\phi}_0'(\omega/\pi)/\pi |  
 \\ &\quad \quad\quad\quad+ | \hat{h}_0(\omega)\hat{\phi}_0''(\omega/\pi)/(\pi)^2| \nonumber
 \\&\leq \frac{\pi}{8}\cdot C_{\phi_0}+\frac{C_{\phi_0'}}{\pi}+\frac{\pi}{2}\cdot \frac{C_{{\phi}_0''}}{(\pi)^2},\nonumber
 \end{align}
 where $C_{\phi_0''} = \sup\limits_\omega |\hat\phi''_0(\omega)| $.
Substituting the above bounds on $|\hat{h}'(\omega)|$ and $|\hat{h}''(\omega)|$ into \eqref{eq:term2} and then combining the result with  \eqref{eq:term1} and \eqref{eq:split} yields the desired result on $\|h\|_{L_1}$. The statement on $\|h_r\|_{L_1}$ follows by observing that $h_r$ is the convolution of $h$ with itself $r$ times. As $h$ is in $L_1$, $\|h_r\|_{L_1}\leq\|h\|_{L_1}^r$. \qed

\section*{Acknowledgment}
The authors would like to thank Mark Iwen for useful discussions.


%

\bibliographystyle{IEEEtran}
\bibliography{quantization}

\end{document}